\documentstyle[prl,aps,preprint]{revtex}

\tightenlines
\begin{document}

\draft
\title{Self-organization, resources and strategies in a minority game}
\author{Horacio Ceva$^*$}
\address{Departamento de F\'{\i}sica, Comisi\'on Nacional
de Energ\'{\i}a At\'omica,\\
Avda. del Libertador 8250, 1429 Buenos Aires, Argentina}
\date{24 September, 1999}
\maketitle

\begin{abstract}
We find that the existence of self-organization of the members  of a
recently proposed minority game, depends on the type of update
rules used. The resulting resource distribution is studied in some detail,
and a related strategy scheme is considered, as a tool to improve the
understanding of the model.
\end{abstract}

\pacs{PACS numbers: 05.65.+b, 02.50.Le, 64.75.+g, 87.23.Ge}

\vskip2pc]


The emergency of organization inside a population can be the result of local
interactions between its members. This type of problems have been under
study for a long time, and can be schematically reduced, for instance, to an
Ising-like model. A problem that has recently become of interest is the self
organization of a population {\it without} direct interactions between its
members, but with a feedback mechanism related with its {\it collective}
behavior. The minority game, introduced by Challet and Zhang \cite{zhang},
addresses one of the simplest situations of this kind. In this model every
member of a population has to choose from a simple alternative, without
knowing what the other members will do. Simple cases are: to buy or sell in
a stock market, to select one of two possible routes, etc. At the end of the
day, the winners are those `agents' that happen to be in the minority side.
Feedback is established by a reward system for winners and losers. In more
general terms, these problems are nothing but simple examples of a situation
where there is a competition for a limited resource (money, food, free
highways, etc.), and individual members of a population adapt their behavior
following their (recent) experiences. Arthur was the first to propose this
type of approach\cite{arthur}, in what now is known as the {\it El Farol}
bar problem.

The specific form in which every member of the population makes his choise
is generically designated as his `strategy'. Different versions of the model
are characterized by \ this strategy selection. In this work we will address
the model proposed by Johnson {\it et al.} \cite{johnson}. As in all
minority games, there is an odd number of agents $N$, every one choosing
between option ``0'' (e.g. to buy an asset) and option ``1'' (to sell the
asset). After all agents have made their choise, the winners, {\it i.e.}
those in the minority group, gain a point,\ while those in the mayority
group lose a point. A single binary digit, $0$ or $1$, signals the winner
option. Each agent knows beforehand the previous $m$ outcomes of the game,
as well as the outcomes of the most recent occurrences (`histories') of all $%
2^{m}$ possible bit strings of length $m.$ Now, Johnson {\it et al.} assign
to each agent a single number $p$ ( $0\leq p\leq 1)$: given a history, the
agent will either choose the same outcome as that stored in the memory, with
probability $p$, or will choose the opposite with probability ($1-p$).
Strategies can be modified, following the evolution of the game. Thus, if an
agent's account is below a threshold value $d<0$ , he gets a new strategy,
whose value $p^{\prime }$ is chosen with an equal probability from the
interval $(p-r/2,p+r/2)$, where $0\leq r\leq 2$; in what follows we will use
the simpler notation $p\rightarrow p^{\prime }=p\pm \Delta p$.
Simultaneously (and to some extent, arbitrarily), his account is reset to
zero. As we will discuss below, the existence of negative points, combined
with the behavior at the threshold, introduce some confusion at the time of
considering the resources. In the following, we will refer to this
combinations as the $d$-rule.

The work of Johnson {\it et al}. has shown that, as a result of the
correlations established by those rules, agents self-organize, in such a way
that the frequency distribution $P(p)$ becomes strongly peaked around both $%
p\simeq 0$ and $p\simeq 1$ (see curve $R1$ in Fig. \ref{fig1}).

Interesting as it is, this work leaves open some questions. First, it would
be interesting to check the robustness of self-organization, under changes
in the strategy actualization rules. As we will see, it also is of interest
in this case to study with some detail the question of the resulting
distribution of the resources.

Our notation is as follows. A single realization of the game involves $n_t$
time steps. All results are then averaged over $n_s$ different samples. The
total amount of points to be distributed in this process is $T=N*n_t*n_s$.
We call $n_{+}$ ($n_{-}$) the number of positive (negative) points, $i.e.$
those assigned to winner (loser) agents. Obviously, $T=n_{+}+n_{-}$. There
is also certain amount of points, $N_{lost}$, that are eliminated from the
game, namely those assigned to any agent changing his strategy, $%
p\rightarrow p^{\prime }$. After all $n_s$ games are played, there will be $%
N_{acc}$ accumulated points, resulting simply from the sum of all accounts,
at the end of every game. Note that, in general, $T>N_{lost}+N_{acc}$,
because there are both positive and negative contributions to the accounts.
Whenever necessary, we used reflective boundary conditions.

We have made extensive numerical simulations, both with $p\rightarrow
p^{\prime }=p\pm \Delta p$, the original rule ($\equiv R1$), and also with $%
p\rightarrow p^{\prime }=(1-p)\pm \Delta p$ (rule $R2$), a seemingly minor
modification of the actualization rule for the strategies. In the latter
case, any loosing agent will change his mind and pick a `complementary'
strategy; in other words, if the initial selection was, say, to choose
preferentially option ``0'' then, after losing, the agent will rather prefer
option ``1''. The resulting distribution functions are shown in Fig.\ref
{fig1}. Our results for $R1$ reproduce (without noise) those of Johnson {\it %
et al.} It is apparent that there are important differences between both
cases:\ while self-organization shows up very clearly for $R1$, it is
practically absent in $R2$, which remains near to its initial (homogeneous)
distribution \cite{distribution}. This result shows, apparently, that the
presence of self-organization itself depends on the kind of strategy
employed.

Consider now the question of the distribution of available resources ({\it \
i.e.} points). We have already mentioned that every agent losing more than $%
d $ points gets his account reset to zero, whereby all those points leave
the game. This introduces some confusion at the time to interpret our
results. The standard interpretation \cite{zhang} only takes into account
positive points. I f we ignore for a moment the $d$-rule, ans simply
consider all positive points added, we found that the accumulated earning
per time step, $n_{+}/T,$ is $\approx 0.47$ in both cases, within a small
error. Notice that this is very near $(N-1)/2N=0.495$, the maximum possible
gain with $N=101$. The behavior of both cases, however, is also very
different in this regard. In fact, the distribution of positive points
earned by an agent, $C^{+}(p)$, follows closely the form of the
corresponding $P(p)$ shown in Fig.\ref{fig1},{\it \ i.e. }while earning is
concentrated around the extrema for $R1$, it is distributed for $R2.$

On the other hand, application of the $d$-rule modifies sensibly this
interpretation. The number of points earned by an agent with strategy $p$, $%
C(p)$, is the algebraic sum of both positive and negative points. Every time
$C(p)<d$, all these points (positive and negative) are discarded. At the end
of all games, it is natural to choose the ratio $G=N_{acc}/T$ as the
magnitude of interest. We find that $G$ is always positive, but vanishes as $%
T\rightarrow \infty $. In fact, $N_{acc}$ is proportional to both $N$ and $%
n_{s}$, and our results imply $G\sim 1/n_{t}$. Thus, for instance, for $%
n_{s}=500,$ and $n_{t}=10^{6}$, it already is $G\simeq 0.00002$. In other
words, although there is self-organization, as described by the work of
Johnson{\it \ et al}., the net resources distributed between all agents are
vanishingly small. It is worth mentioning that this is {\it not} the result
of adding negative and positive accounts: in the mean they are mostly
positive. Simply enough, $C(p)$ follows closely the behavior of $P(p)$, but
its magnitude vanishes in this limit. A good amount of the points involved
in the game turn out to be in $N_{lost}$ and, what is more important, $%
N_{lost}\simeq n_{-}-n_{+}$. This can be described by telling that in this
case {\it there are no winners in the game }\cite{rule}. This situation is
still more pronounced in case one uses $R2$, the alternative strategy rule.

In fact, it is only because the accounts are adjusted periodically to zero,
that they appear to have mostly positive balance at the end of the games. In
their study of this model, D'hulst and Rodgers \cite{d'hulst} used the
Hamming distance between strategies and concluded that, on average, the
number of points earned by an agent, $C(p,t)$, evolves with time $t$
following

\begin{equation}
C(p,t)=-(\frac 12-\tau (p))\text{ }(t-t_0)  \label{d+r}
\end{equation}

until $C<d$, at which point is set equal to zero. In the above expression $%
\tau (p)<1/2$ and $t_{0}$ are constants. This is a sawtooth function of $t$
that is always negative, or vanishes. This analysis, however, does not takes
into account properly the role of the $d$-rule, as can be seen in the
following example. Consider a game where the winner is always the same
agent, while all others lose. After $L\;$time steps, the winner agent will
have $L\;$ points, while the remaining $(N-1)$ agents will have $-L$ each.
Whenever $-L<d$, the $d$-rule implies that only the winner keeps his points.
Therefore, after a while, the net amount of points of all players is
necessarily positive.

We now turn our attention to a related type of strategy. Our main concern
here is to understand how we can improve the resource distribution, using
rules similar to those of Ref. \cite{johnson}.

Consider a rule $p\rightarrow p^{\prime }$ which is intrinsically asymmetric
(rule $R3$). In this case,

\begin{equation}
p^{\prime }=p_0\pm \Delta p  \label{distrib}
\end{equation}

where $0\leq p_0$ $\leq 1$ is constant.

Application of Eq.\ref{distrib} will move agents to the neihborhood of $%
p_{0} $. Eventually, however, there will be a majority of agents in this
place, and therefore all others players will win, establishing a stationary
state. We want to know the dependence of $G$\ on $p_{0}$, $\Delta p$ and $r$%
. Note that it is possible to describe some cases $p\rightarrow p^{\prime }$
as a superposition of situations with this type of update scheme. Figure \ref
{fig2}$(a)$ illustrates the case $p_{0}=0.8$, $r=0.2$. The resulting
frequency distribution is asymmetric. In this case {\it there are winners},
namely those agents that manage to have their strategy below 0.5 The left
side of Fig.\ref{fig2}$(a)$ shows the gain as a function of $p$. In Fig.\ref
{fig2}$(b)$, on the other hand, we have the gain $G$ ({\it i.e. }the
integral of that shown in $(a)$) as a function of $p_{0},$ for a fixed value
of\ $r$.

ALso, and rather unexpectectly, we can see in Fig.\ref{fig3} that $G$ is
almost independent of $r$, until it approaches $r\approx 1$, where there is
something analogous to a `phase transition'; it probably corresponds to the
`transition' between localization around $p_{0}$, and delocalization. It
should be emphasized that these results are associated with the use of the $%
d $-rule. Within the standard interpretation, it can be seen that the gain
increases with $r$, at least for $r$ smaller than $\approx 1$.

Finally, it should be pointed out that, although the present version shares
the main ideas of the original minority game\cite{zhang}, in some respects it
does not follows the same behavior.   Recent work  \cite{challet} study the
model of Callet and Zhang in terms of the variable $\alpha =P/N$ (in our
case, $P=2^{m}$), and the variance of the time series, $\sigma
^{2}=<(n_{-}-n_{+})^{2}>$ or, more especifically, the reduced variance $%
z\equiv \sigma ^{2}/N$. As it is well known, the random agent case \ is
given by $\sigma ^{2}=N,$ ${\it i.e.}$  $z=1.$ This value is attained for $%
\alpha _{r}\approx 0.2$. Smaller values of $\alpha $ produce a
worst-than-random answer ($z>1$), while the game output improves if $\alpha
>\alpha _{r}$\ ($z<1$). Moreover, they have identified two `phases',
characterized by the behavior of $z=z(\alpha ).$\ For  $\alpha <\alpha _{c}$%
, the reduced variance decreases with $\alpha $, but for $\alpha >\alpha _{c}
$ it becomes an increasing function of $\alpha .$ Numerical simulations give
the critical value $\alpha _{c}\approx 0.34$ A theoretical description of
this game has been developed \cite{challet}, based on an analogy with spin
glasses.\ In any case, it is apparent that, for fixed $N$, the response of
the game is strongly dependent with $P\;$({\it i.e.} $m$ in our case). We
have not completed a systematic study in this respect, but it is rather
clear that the present model is, in fact, almost totally independent of the actual
value of $m$. On the other hand, we find $z\approx 0.04-0.08$ for our three
upgrade rules (although $\alpha =2^{3}/101\simeq 0.08$). This illustrates an
important difference between both formulations.

This work was partially supported by EC Grant ARG/B7-3011/94/27, Contract
931005 AR.

\vspace{0.5cm}

($^*$)E-mail address: ceva@cnea.gov.ar

\begin{figure}[tbp]
\caption{Frequency distribution function for two different strategy
actualization rules. {\it R1}: $p\rightarrow p^{\prime }=p\pm \Delta p$;
{\it R2}: $p\rightarrow p^{\prime }=(1-p)\pm \Delta p$ . In both cases $%
N=101 $, $n_t=10^5$, $n_s=10^4$, $d=-4$, $m=3$, $r=0.2$.}
\label{fig1}
\end{figure}

\begin{figure}[tbp]
\caption{Strategy rule $p\rightarrow p^{\prime }=p_0\pm \Delta p$, $N=101$, $%
n_t=10^5$, $n_s=10^4$, $r=0.2$, $d=-4$, $m=3$. The line in ($b$) is only a
guide to the eyes}
\label{fig2}
\end{figure}

\begin{figure}[tbp]
\caption{Strategy rule $p\rightarrow p^{\prime }=p_0\pm \Delta p$. $n_t=10^5$%
, $n_s=500$, $N, d, m$ have the same values as in Fig.\ref{fig2}. Continuous
lines are only a guide to the eyes.}
\label{fig3}
\end{figure}

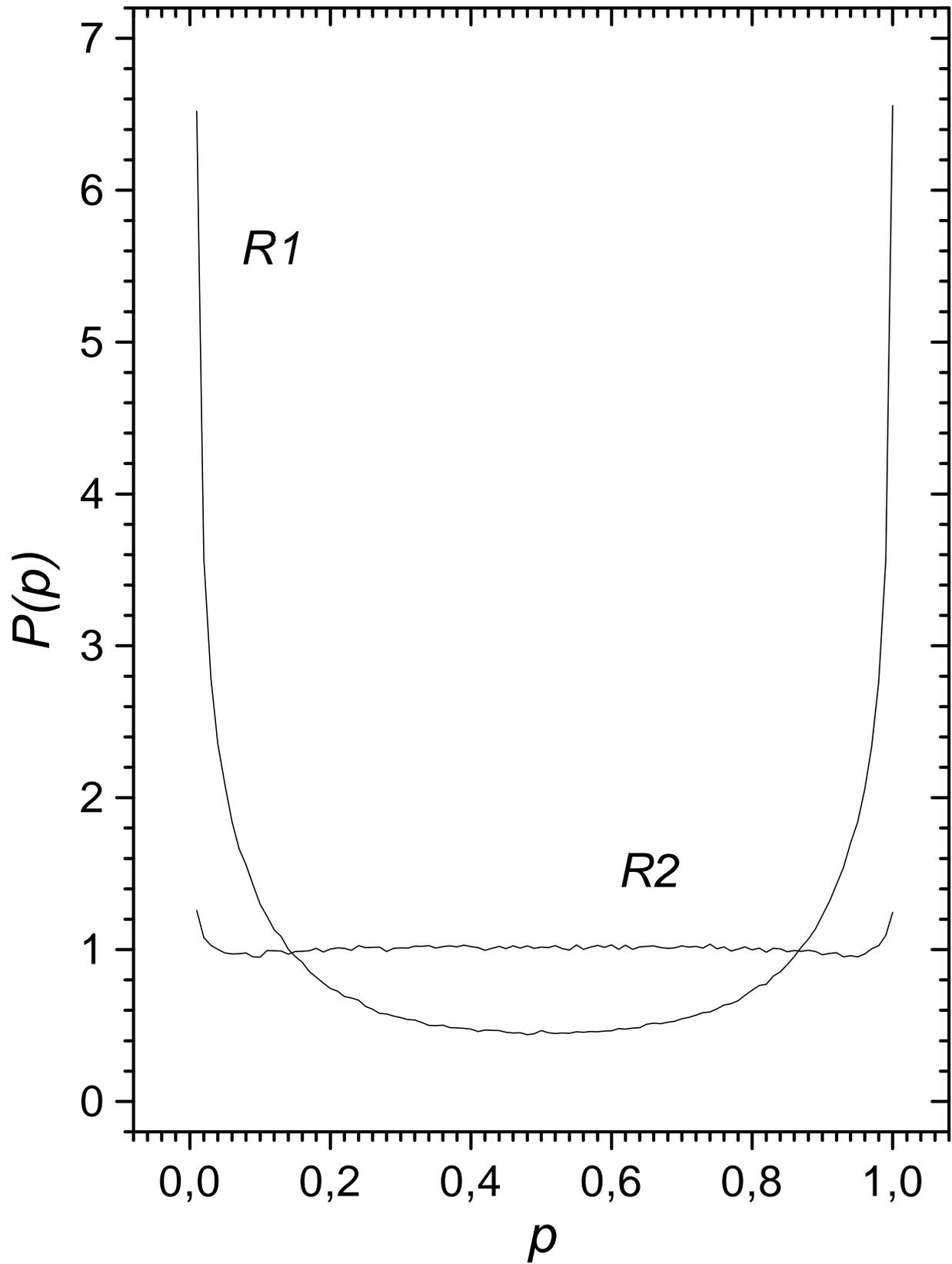
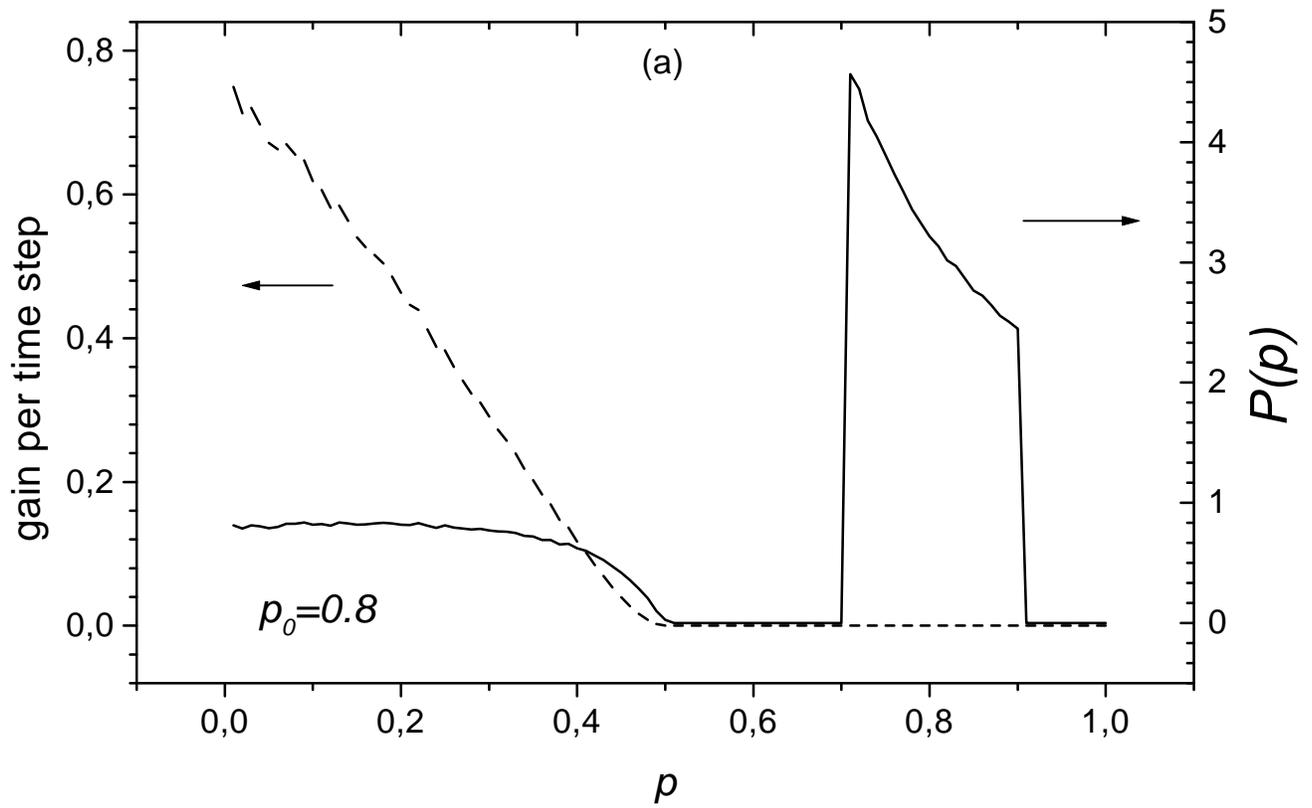
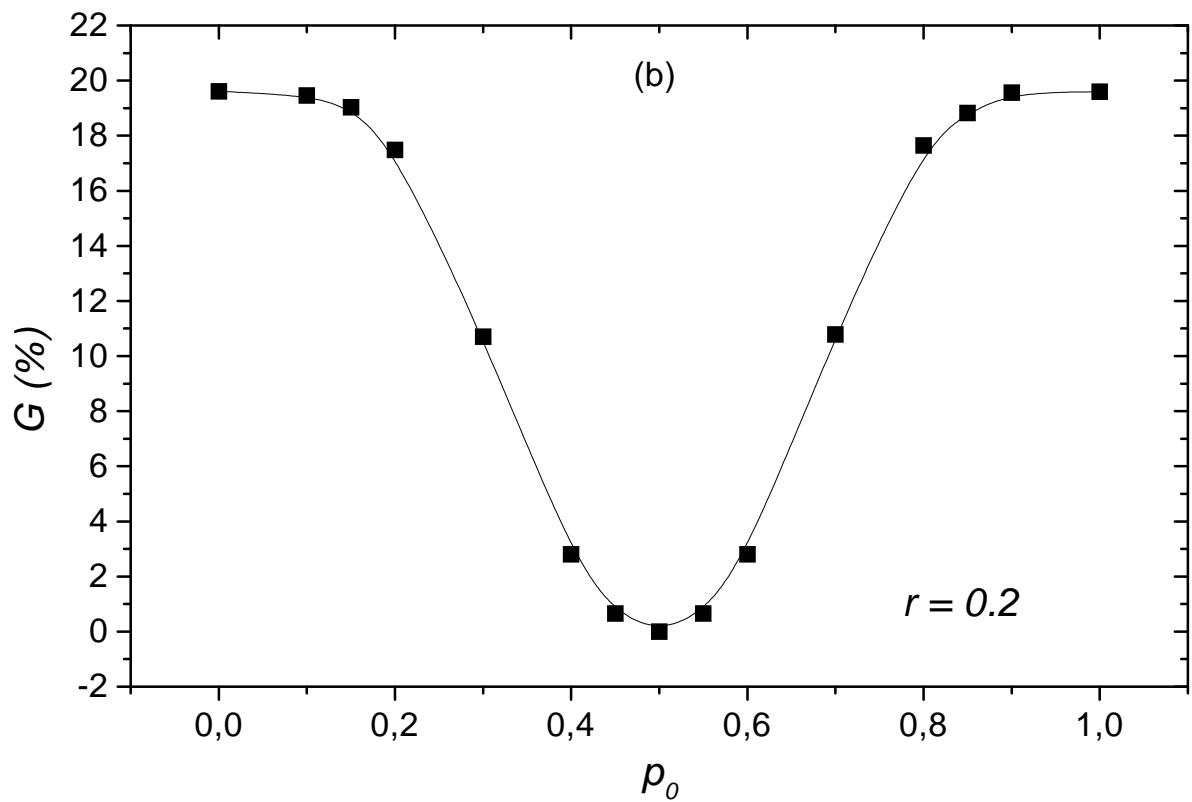
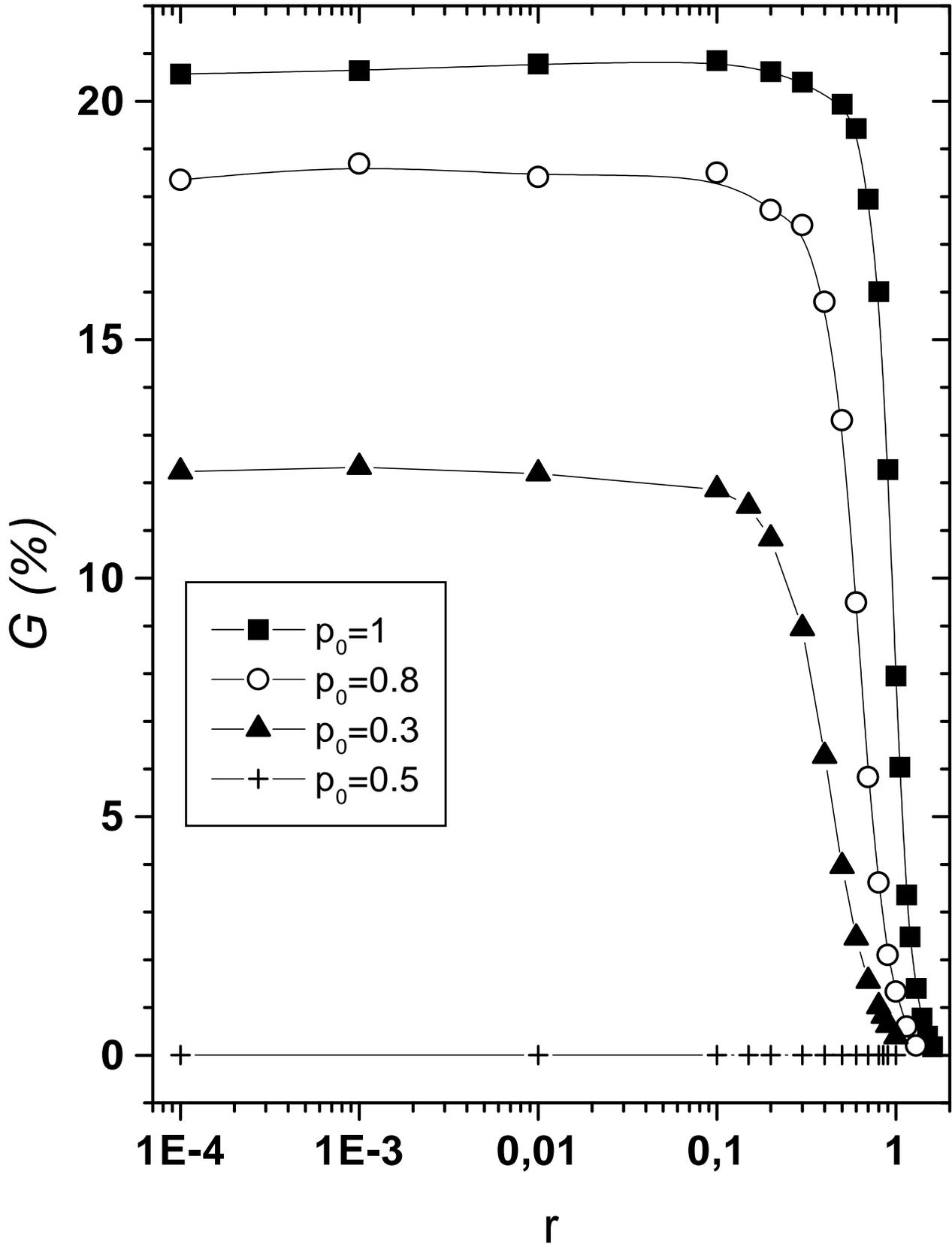
\end{document}